\documentstyle[aps,prl,twocolumn,psfig,floats]{revtex}
\begin{document}
   \preprint{\vbox{\hbox{UW-PT-01/15}},
   \vbox{LBNL-44551}}                                     %
   \draft                                                                      %
   \wideabs{                                                                   %
   \title{Unification without Unification}    
   \author{Neal Weiner}                    %
   \address{                                                                   %
   Department of Physics,                                                      %
   University of Washington,                                                   %
   Seattle, WA~~98195, USA}                                                   
   \date{\today}                                                               %
   \maketitle                                                                  %
   \setcounter{footnote}{0}                                                    %
   \setcounter{page}{1}                                                        %
   \setcounter{section}{0}                                                     %
   \setcounter{subsection}{0}                                                  %
   \setcounter{subsubsection}{0}                                               %
   \begin{abstract}
   	The logarithmic running of the gauge couplings $\alpha_{1}$, 
   	$\alpha_{2}$ and $\alpha_{3}$, indicates that they may 
	unify at some 
   	scale $M_{GUT} \sim 10^{16} {\rm GeV}$. This is often taken to imply that the 
   	standard model gauge group is embedded into some larger simple 
   	group in which quarks and leptons are placed in the same 
   	multiplet. These models have generic features, such as proton 
   	decay, and generic problems, namely the splitting of the Higgs 
   	doublet and triplet. Inspired by the recent discussion of dimensional 
   	deconstruction, we propose an interesting alternative: 
   	we postulate a strongly coupled $SU(3)\otimes SU(2) \otimes U(1)$, 
	which is not the remnant of a GUT, and is  
   	Higgsed with a weakly coupled $SU(3)\otimes SU(2) \otimes U(1)$, 
   	which is the remnant of a GUT, or with a GUT group directly, 
	into the diagonal subgroup. In this ``collapsed GUT'' mechanism,
	unification of coupling constants in the 
	low energy theory is expected, but proton decay and the 
	doublet/triplet splitting problem are entirely absent.
   \end{abstract}
}
%
\newcommand{\gsim}{ \mathop{}_{\textstyle \sim}^{\textstyle >} }
\newcommand{\lsim}{ \mathop{}_{\textstyle \sim}^{\textstyle <} }
\newcommand{\vev}[1]{ \left\langle {#1} \right\rangle }
\newcommand{\bra}[1]{ \langle {#1} | }
\newcommand{\ket}[1]{ | {#1} \rangle }
\newcommand{\ev}{ {\rm eV}}
\newcommand{\kev}{ {\rm keV}}
\newcommand{\mev}{ {\rm MeV}}
\newcommand{\gev}{ {\rm GeV}}
\newcommand{\tev}{ {\rm TeV} }
\newcommand{\mpl}{$M_{Pl}$}
\newcommand{\mw}{$M_{W}$}                                                            %
%
\vskip 0.3in                                                                %
\section{Introduction}
The standard model, consisting of the gauge group $SU(3)\otimes SU(2) 
\otimes U(1)$ broken at the weak scale to $SU(3)\otimes U(1)$, has 
been extremely successful. Nonetheless, it has a number distasteful features, 
which has prompted a great deal of study 
into the possibilities of physics beyond the standard model.

It is exceptionally notable that two significant proposals of physics 
beyond the standard model seem quite complementary, namely Grand 
Unified Theories (GUTs) and supersymmetry (SUSY). As precision 
measurements on the gauge couplings $\alpha_{1}$, $\alpha_{2}$ and 
$\alpha_{3}$ have improved, it has become increasingly clear that the 
original GUTs - without supersymmetry \cite{Georgi:1974sy,Georgi:1974yf} 
- do not 
agree with the measured value of $\sin^{2}\theta_{W}$. However, GUTs 
with supersymmetry \cite{Dimopoulos:1981zb,Sakai:1981gr} 
seem to agree quite well with precision data 
\cite{Dimopoulos:1981yj}. 
Indeed, this is often considered a great success of supersymmetry.

A generic feature of GUTs is the instability of the proton, which 
occurs dominantly through either X and Y boson mediated dimension six 
operators, or through triplet Higgsino mediated dimension five 
operators. The triplet Higgs mass is a free parameter, but often 
the dimension five operators are the dominant source of proton decay 
\cite{Sakai:1982pk,Weinberg:1982wj}. 
Together with precision measurements on the gauge couplings, 
this has been used to exclude various models of grand unification 
\cite{Hisano:1993jj,Goto:1999qg}.

While GUTs are theoretically appealing, they are not without problems. 
Probably the greatest is the doublet/triplet splitting problem. 
Because of the larger gauge group, the Higgs comes with a triplet 
partner, whose mass must in general be near the GUT scale. 
In ``minimal'' $SU(5)$, the GUT is broken by 
a field $\Sigma$, which transforms as a ${\bf 24}$ under $SU(5)$. 
The bare mass superpotential term $m H \overline H$ is then tuned 
against a $\Sigma H \overline H$ term to give the doublet a small 
mass, while leaving the triplet with a mass $O(M_{GUT})$. 

A number of solutions exist for this, including the missing partner 
mechanism \cite{Masiero:1982fe,Grinstein:1982um}, 
the Higgs as a pseudo-goldstone boson 
\cite{Inoue:1986cw,Barbieri:1993yy}, 
and others. We shall not discuss the merits and drawbacks 
of each here, but clearly some solution is in order.

\subsection{Grand Unification?}
Before we continue further, let us reexamine what the evidence is 
for grand unification. Given the particle content of the standard 
model, we can study the renormalization group evolution of the gauge 
couplings from the weak scale to higher energy scales. We then 
extrapolate over {\em fourteen decades of energy}, assuming nothing 
but MSSM fields enter into the RGEs. This evidence for grand 
unification is quite indirect.

At the same time, there is also indirect 
evidence against grand unification. There is 
the absence of any proton decay signal, but additionally, the 
expected relations between $m_{e},m_{\mu}, m_{d}$ and $m_{s}$ fail by 
an order of magnitude. Given the additional complexities that are 
necessary to solve the doublet/triplet splitting problem, it is 
perhaps worthwhile to question whether we {\em must} read the gauge 
coupling unification as an indication of the standard model being 
embedded in a unified group. Put simply: can we understand coupling 
unification without a conventional GUT?

Various proposals have put forth to this end. For instance, in 
\cite{Cabibbo:1982hy} it was proposed that coupling constant unification 
could occur in a strongly coupled theory. Other 
possibilities include using a different group structure \cite{Riazuddin:1986jy}, or with 
unification at the string scale \cite{Kaplunovsky:1992vs}, 
without a grand unified group.

In this letter, we will see how an enlarged gauge symmetry can 
naturally give gauge coupling unification without having a grand 
unified theory in the conventional sense. The outline is as follows: 
in section \ref{sec:general} we shall discuss how Higgsing a strongly 
coupled sector into a weakly coupled remnant of a grand unified group 
gives the appearance of unification, what we lose from such a 
scenario, and how such breaking might occur. In section 
\ref{sec:lowscale} we comment on such a scenario in theories with 
\tev-sized extra dimensions and gauge coupling unification at a low 
scale, $O(10\tev)$.

\section{Unification without Unification}
\label{sec:general}
We begin by considering the well known case of two copies of a single 
gauge group $G$, with couplings $g_{1}$ and $g_{2}$. 
If the theory is Higgsed down to the diagonal subgroup, the gauge 
coupling of the resulting massless gauge boson is given by the 
well-known formula
\begin{equation}
	\frac{1}{g_{eff}^{2}}=\frac{1}{g_{1}^{2}}+\frac{1}{g_{2}^{2}}.
\end{equation}
The situation we shall be most interested in is the case in which 
$g_{1}$ is small and $g_{2}$ is large. In this case, $g_{eff} \approx 
g_{1}$.

With this simple fact in hand, we can consider the following scenario. 
Consider a model in which the gauge symmetry is $G_{W}\otimes G_{S}$. 
As before, $G_{W}$ will be weakly coupled, while $G_{S}$ will be 
strongly coupled at the GUT scale. $G_{W}$ will be some semi-simple group which 
contains $SU(3)\otimes SU(2) \otimes U(1)$ as a subgroup, but not the 
groups under which quarks and leptons are charged. Instead, let us 
take $G_{S}$ to also contain a copy of $SU(3)\otimes SU(2) \otimes 
U(1)$ - although it may be larger -
under which quarks and leptons are charged. 

At some scale $M_{D}\le M_{GUT}$, we assume some additional dynamics acts 
to Higgs the $G_{W}\otimes G_{S}$ down to its diagonal
$(SU(3)\otimes SU(2) \otimes U(1))^{2}$ 
subgroup. This may occur simultaneously with the GUT breaking 
or at a lower scale. For the practical purpose of achieving gauge 
coupling unification, it is generally best that these happen 
simultaneously.
However, since $g_{S} \gg g_{W}$, then as before we have 
$g_{eff}\approx g_{W}$, except now {\em the standard model fields are 
charged under this group!} 

Let us study the RG evolution of the gauge couplings to extrapolate to low 
energies. At the GUT scale, the 3-2-1 gauge couplings are strong, 
while the $SU(5)$ coupling is weak. Since we are assuming that the 
only non-$SU(5)$ complete multiplets lie in the standard model 
sector, the gauge couplings run from the GUT scale as
\begin{eqnarray}
	\alpha_{i}^{-1}(\mu) &=& \alpha_{i}^{-1}(M_{GUT})\\ &&+ ({\rm MSSM \>running}) + 
	({\rm universal\> running}),\\
	\alpha_{5}^{-1}(\mu) &=& \alpha_{5}^{-1}(M_{GUT})\\&&+ ({\rm 
	universal\> running}).
\end{eqnarray}
When we Higgs to the diagonal subgroup, leaving only 3-2-1, the gauge 
couplings of the remaing gauge group are just the sum of these. 
Explicitly, at a scale $\mu< M_{D}$, we have
\begin{eqnarray}
	\alpha_{(LE)i}^{-1}(\mu) &=& \alpha_{i}^{-1}(M_{GUT}) 
	+\alpha_{5}(M_{GUT})\\&&+\frac{b_{i,MSSM}}{2 \pi} \log(M_{GUT}/\mu) \\
	&&+ ({\rm SU(5)\> universal}),
\end{eqnarray}
where we use LE (low energy) to distinguish the gauge coupling of the 
remaining massless group after Higgsing from the corresponding 
coupling above $M_{D}$. Note the MSSM running is independent of the 
scale of breaking. 

At the weak scale, 
this will be indistinguishable from an ordinary GUT up to corrections 
arising from the $\alpha_{i}^{-1}(M_{GUT})$.
We would like these corrections to be comparable to those expected in 
ordinary GUT theories, requiring $\alpha_{i}(M_{GUT}) \ge 1$, so the theory is 
somewhat strongly coupled, but still perturbative.

We now have a remarkable situation: at low energies, the gauge 
couplings are consistent with being embedded within a grand unified group. 
However, there is no proton decay from X and Y exchange as quarks and 
leptons are not charged under the GUT. There is no proton decay from 
the Higgs triplet because there is no Higgs triplet in the theory. 
Moreover, the Yukawas will not obey any GUT relationships.

This scenario is reminiscent of a model of doublet/triplet splitting 
proposed in \cite{Hotta:1996cd}, in which the gauge group $SU(5)\otimes 
SU(3)\otimes U(1)$ was 
postulated to give the Higgs triplet a large mass. The differences 
are profound, however: there, the standard model really was embedded 
into a unified group. Here it is not.

\subsection{Collapsed GUTs and Dimensional Deconstruction}
The recent surge of interest in ``Moose'' \cite{Georgi:1986hf} or ``Quiver'' 
\cite{Douglas:1996sw} models has been spurred by the realization that such 
theories can serve as UV completions of higher dimensional gauge theories 
\cite{Arkani-Hamed:2001ca,Hill:2000mu}, 
and provide useful features of the higher dimensional 
theories without the associated UV problems.

This mechanism also has a clear analog in higher dimensions: that of 
the GUT broken by orbifold boundary conditions \cite{Nima}. The idea of 
breaking grand unified theories by Wilson lines has existed for a 
great while \cite{Witten:1985xc,Breit:1985ud,Sen:1985af}, 
and has recently seen a 
resurgence of its application in realistic model building 
\cite{Kawamura:2001ev,Kawamura:2001ir,Altarelli:2001qj,Hall:2001pg}. 
In these models, at a specific 
point in the fifth dimension, the SU(5) gauge transformation 
vanishes. We can identify this point as the site on a Moose diagram 
where we merely have 3-2-1 gauge group. While a direct deconstruction would in 
general have multiple SU(5) sites, while we have but one, the connection 
is clear. A more general discussion of deconstruction would be 
interesting \cite{GC}.

\subsection{What have we lost?}
Grand unified theories do have many desirable features \cite{pati}.
Now that the standard model is not grand unified, we lose many of these, 
but not as much as might be expected. 

For instance, we have the charge assignments of the MSSM chiral 
matter fields. Since any underlying theory can only should only 
generate consistent quantum theories, we 
can still understand this through anomaly cancellation. Nonetheless, 
ther overall normalization of hypercharge is still undetermined. One 
might make assumptions that the fundamental string theory naturally 
gives the proper normalization, but this is not necessary. A perhaps 
more natural assumption is that 3-2-1 is contained in a nonabelian 
gauge group, in which the charge assignments are automatic. Two 
obvious examples would be $SU(3)^{3}$\cite{Glashow:1984gc}
and $SU(4)\otimes SU(2) \otimes SU(2)$ \cite{Pati:1973uk}.

Although there is no right handed neutrino in this model,
we still expect 
heavy states at $M_{GUT}$, so if lepton number is broken there, we 
still understand neutrino masses. In any event, there are a number of 
ways to understand neutrino masses in supersymmetric theories 
\cite{Hall:1984id,Arkani-Hamed:2000kj,Arkani-Hamed:2000bq,Borzumati:2000mc}.

Additional symmetries are easily added to the theory, such as lepton 
number, $B-L$, and Froggatt-Nielsen symmetries. The unification of 
bottom and tau Yukawas is an important success in certain regions of 
parameter space, but it is not a 
generic success of the MSSM \cite{Polonsky:1996zc}. 

In conclusion, while these are many successes which arise from 
grand unification, for the most part they can be included in this 
framework.

\subsection{Breaking to the diagonal subgroup}
It is simple to break $SU(5)\otimes SU(3) \otimes SU(2) \otimes U(1)$
to the diagonal subgroup. An explicit 
linear sigma model was given in \cite{Cheng:2001an}.
One could also imagine using strong dynamics if fields charged under 
$SU(5)$ and under $SU(3)\otimes SU(2) \otimes U(1)$ condensed, 
breaking to the diagonal.
Of course, this has swept various questions into the guise of strong 
dynamics. For instace,
we have no clear understanding of why it is broken precisely 
to the diagonal subgroup, rather than some other subgroup. However, 
for a non-SUSY GUT model, these are  
certainly necessary questions, and warrant the development of a 
realistic model.

In general, the $3-2-1$ sector should be strongly coupled at or near 
the GUT scale. Absent additional matter content, 
$SU(3)$ remains asymptotically free. 
Thus, for $M_{D}$ to lie significantly below the GUT scale, additional 
multiplets are needed. However, the matter used to break to the 
diagonal subgroup can serve this purpose, for example the fields in 
 \cite{Cheng:2001an}. We also assume 
the fields in the breaking sector should only contain fields 
which appear in complete $SU(5)$ multiplets in order not to spoil the 
quantitative success of grand unification. One can imagine including 
fields in which there were incomplete multiplets, but this can only be 
addressed within a specific model.

Finally, we should make one comment regarding scales: without 
additional matter 
content, $SU(2)_{S}\otimes U(1)_{S}$ is infrared 
free, while $SU(3)$ is asymptotically free. Thus, to have all 
couplings be simultaneously strong at the GUT scale
is a significant constraint on the 
theory. In particular, a strong U(1) at the GUT scale will have a 
Landau pole before the Planck scale, so it cannot be considered truly 
fundamental unless the string scale is lowered.
We should note, however, that these problems are irrelevant once we 
have embedded the U(1) into a non-Abelian group. Since this is already 
motivated by hypercharge assignment, the asymptotic freedom of the model 
is not a great concern.

\section{TeV scales and phenomenology}
\label{sec:lowscale}
Does this scenario have any unique phenomenology? Outside of the 
absence of proton decay and the non-unification of Yukawas, there is 
no obvious signal. However, there is a great deal of dependence on 
the scale at which the strong and weak groups are Higgsed to the 
diagonal subgroup. If this occurs at a scale significantly 
below $M_{GUT}$, there could be noticeable threshold effects if any 
multiplets are split. Of course, we already expect some level of 
non-universal effects from the $3-2-1$ gauge couplings themselves.

Other possibilities arise when we add additional structure.
In supersymmetric theories, the RG contribution to the soft scalar 
masses should be modified above $M_{D}$, and so could be 
incompatible with mSUGRA depending on the size of the effect.
Moreover, because the gauginos will mix, their masses need not unify.

We now have the possibility of adding other gauge groups, such as 
$U(1)_{B}$ which are incompatible with $SU(5)$, for instance, so long 
as it is made anomaly free. There are no doubt other interesting 
extensions. 

There is another exciting possibility, however, which is that the 
Higgsing to the diagonal subgroup occurs near the TeV scale. In such a 
scenario, at upcoming colliders, we would expect to see new $3-2-1$ 
gauge bosons with strong couplings to quarks and leptons. Such a 
possibility is especially attractive in models with \tev scale GUTs 
proposed in ref. \cite{Dienes:1998vh,Dienes:1999vg}. 
Quantitatively, we must address such a possibility. If there are \tev 
scale GUTs, then quantitative unification is not as precise in SUSY 
GUTs, so the constraint on the strength of the gauge couplings at the 
GUT scale may be weaker. 

\section{Conclusions}
The apparent unification of gauge couplings has given us motivation 
for considering grand unified theories. However, the absence of 
proton decay, the non-unification of Yukawas and the doublet/triplet 
splitting problem warrant consideration of other possibilities. At 
the most minimal level, we only have the unification of coupling 
constants, so we need also ask whether that alone can be explained 
without conventional grand unification.

Here we have demonstrated a scenario in which this can happen naturally.
By Higgsing a weakly coupled $SU(3)\otimes SU(2) \otimes U(1)$ 
arising out of a unified group with a 
strongly coupled copy under which quarks and leptons transform, it 
will be in many cases indistinguishable from a unified theory, up to 
the absence of proton decay and constraints on the Yukawas. At 
present, we have merely described a mechanism, involving unknown 
dynamics. The development of a complete model is an worthwhile task.

Such a scenario might be interesting to consider if Higgsed to the 
diagonal group at the TeV scale, or when embedded into TeV scale GUTs. 
There is a wealth of phenomenology to be undertaken.

\vskip 0.25 in
{\noindent \large \bf Acknowledgments}
\\
We would like to acknowledge David E. Kaplan and Yasunori Nomura for 
useful discussions, and to Ann Nelson for reviewing the manuscript 
and useful discussions. We would also like to thank the organizers of 
the visitor program at Fermilab where this work was completed. 
This work was partially supported by the DOE
under contract DE-FGO3-96-ER40956. 

\setcounter{footnote}{0}

\end{document}